\documentclass[%
superscriptaddress,
 amsmath,amssymb,
 aps,
 pra,
floatfix,
twocolumn
]{revtex4-2}
\usepackage[T1]{fontenc}
\usepackage[final]{graphicx}
\usepackage{dcolumn}
\usepackage{amsfonts}
\usepackage{amssymb}
\usepackage{pbox}
\usepackage{amsmath}
\usepackage{amsthm}
\usepackage{mathtools}
\usepackage[normalem]{ulem}

\DeclareMathOperator{\erf}{erf}

\usepackage{caption}
\usepackage{subcaption}
\usepackage{mathtools}
\usepackage{braket}
\renewcommand\bra[1]{{\langle{#1}|}}
\renewcommand\ket[1]{{|{#1}\rangle}}

\usepackage{chngcntr}
\usepackage{multirow}
\usepackage{tabularx}
\usepackage[labelsep=period]{caption}
\date{}
\usepackage{graphicx}
\usepackage{enumerate}
\usepackage{calc}
\usepackage{indentfirst}
\usepackage{booktabs}
\usepackage{dsfont}
\usepackage{arydshln}
\usepackage{bm}
\usepackage{amsthm}
\usepackage{commath}
\usepackage[dvipsnames]{xcolor}

\theoremstyle{definition}

\usepackage{natbib}
\setcitestyle{square,numbers}
\usepackage{changepage}
\usepackage[colorlinks]{hyperref}

\begin{document}

\author{Tomasz Linowski}
\affiliation{International Centre for Theory of Quantum Technologies, University of Gdansk, 80-308 Gda{\'n}sk, Poland}
\email[Corresponding author: ]{tomasz.linowski@ug.edu.pl}

\author{Konrad Schlichtholz}
\affiliation{International Centre for Theory of Quantum Technologies, University of Gdansk, 80-308 Gda{\'n}sk, Poland}

\author{Giacomo Sorelli}
\affiliation{Fraunhofer IOSB, Ettlingen, Fraunhofer Institute of Optronics,
System Technologies and Image Exploitation, Gutleuthausstr. 1, 76275 Ettlingen, Germany}

\title{Quantum-inspired exoplanet detection in the presence of experimental imperfections}

\date{\today}

\begin{abstract}
Ideal spatial demultiplexing (SPADE) is proven to be a quantum-optimal tool for exoplanet detection, i.e., asymmetric source discrimination. However, recent investigations into the related problems of separation estimation and symmetric source discrimination showed its efficiency to be limited in the presence of noise. In this work, we use analytical tools to scrutinize the practical applicability of SPADE and derive the associated optimal decision strategy for exoplanet detection in the presence of experimental imperfections. On the one hand, we find that the probability of detection of noisy SPADE has the same scaling with planet-star separation and relative brightness as conventional techniques, such as direct imaging and coronagraphs. On the other hand, we prove that, due to a superior scaling coefficient under realistic noise conditions, SPADE remains the most efficient method for practical exoplanet detection in the sub-Rayleigh regime.
\end{abstract}

\maketitle

\captionsetup{justification=centerlast}

\section{Introduction}
Since the discovery of 51 Pegasi b, the first planet orbiting a Sun-like star outside the Solar System thirty years ago~\cite{first_solar-like_exoplanet_Mayor_1995}, which partially motivated the 2019 Nobel Prize in Physics, the search for exoplanets remained a key issue in modern astronomy \cite{Exo_1,Exo_2,Exo_3,Exo_5}. Formally, exoplanet detection is a hypothesis testing problem \cite{Harris:64}: given some data describing an outer solar system (e.g., an image), the goal is to test which of the two hypotheses is more likely: (H0) the data originate from a single light source -- the star -- or (H1) the data originate from two light sources -- the star and an exoplanet. More precisely, due to their rarity, exoplanet detection relies on asymmetric hypothesis testing, in which missing an exoplanet is considered significantly more costly than falsely detecting one \cite{hypothesis_testing_book_Van_Trees_2001,hypothesis_testing_Sorelli_2021}.

Unfortunately, due to the proximity of a typical exoplanet to its home star, the efficiency of conventional imaging techniques based on direct imaging drops significantly for separations smaller than the width of the point spread function, due to a phenomenon known as Rayleigh's curse \cite{resolution_survey_denDekker_1997,Fourier_optics_Goodman_2005,Rayleigh_curse_Paur_2018}. 
The problem 
is further complicated by the fact that the light emitted by the planet is far dimmer than that of the star \cite{Exo_4}. Still, a number of quantum-inspired measurement techniques was developed to overcome the limits of direct imaging \cite{PALM_Betzig_2006,PALM_Hess_2006,STORM_Hell_2007,superoscillations_Smith_2016,superoscillations_Gbur_2019,SPLICE_Tham_2017,SPLICE_Bonsma-Fisher_2019,superresolution_techniques_Hemmer_2012,superresolution_techniques_Liang_2021}. Most notably, spatial demultiplexing (SPADE) \cite{superresolution_Tsang_2016,superresolution_starlight_Tsang_2019} was proven to be quantum-optimal for both symmetric and asymmetric hypothesis testing \cite{Exo_4,naive_test_Shapiro_2018,no_point}, with its enhanced sensitivity in the sub-Rayleigh regime verified experimentally \cite{hypothesis_testing_experiment_Zanforlin_2022,hypothesis_testing_experiment_Santamaria_2024,hypothesis_testing_experiment_Wadood_2024}.

Still, 
the real usefulness of any 
measurement 
can be 
practically established only in the presence of experimental imperfections \cite{metrology_noise_Nichols_2016,noise_Oh_2021,Sorelli_practical_superresolution_2021,superresolution_no_location_Grace_2020,Almeida_misalignment_2021,Schlichtholz_2024_SPADE_DYNAMIC}. 
In the particular case of SPADE, the impact of noise in the form of crosstalk, i.e., the possibility of detecting photons in the incorrect output due to defects in the demultiplexer, static and dynamic misalignment, and other imperfections, has already been shown to 
significantly limit both 
the closely related problem of separation estimation \cite{crosstalk_original_PRL,unbalanced_sources_crosstalk_Linowski_2023,Boucher:20,SPADE_Brown}, as well as 
symmetric hypothesis testing. In fact, in the latter case, a decision strategy which is quantum-optimal under ideal conditions was proven to be 
completely useless in the presence of any non-zero amount of noise \cite{hypothesis_testing_Schlichtholz_2024}. 

In this paper, we analytically dissect the practical performance of SPADE in asymmetric hypothesis testing for exoplanet detection. To start with, we calculate how the performance of SPADE is affected by two of the most common 
associated noise sources: 
the aforementioned crosstalk, 
and dark counts \cite{superresolution_limits_SPADE_Len_2020,imaging_noisy_Lupo_2020}, i.e., the possibility of detecting photons originating from sources other than the probed system (typically induced by thermal noise in the detector). Then, we compare the resulting performance with direct imaging, as well as coronagraphs, a family of more refined conventional imaging methods prominent in contemporary astronomy \cite{coronagraphs_Soummer_2003,coronagraphs_Guyon_2006,coronagraphs_review_2018,coronagraphs_NASA_1_Kasdin_2023}. Finally, we provide the toolbox for optimal SPADE-based test for exoplanet detection in the presence of noise: beginning with an explicit derivation of the decision strategy, which differs significantly from the noiseless case, and ending with the analysis of its efficiency for realistic values of planet-star separation, relative brightness, and noise strength, including a specific example inspired by 51 Pegasi b. Our results show that while the impact of noise on SPADE is significant, it remains the closest of the considered exoplanet detection methods to the quantum-optimal bound under realistic conditions. 

The paper is organized as follows. In Sec. \ref{sec:preliminaries}, we describe the hypotheses and the measurement setting. In Sec. \ref{sec:testing_optimal}, we summarize the current state-of-the-art in hypothesis testing for exoplanets. In Secs. \ref{sec:testing_practical} and \ref{sec:coronagraphs}, we analyze the impact of experimental noise on SPADE and compare it with coronagraphs. In Secs. \ref{sec:test_explicit} and \ref{sec:Pegasi}, we derive and discuss the best possible decision strategy for noisy SPADE and subsequently apply it to 51 Pegasi b-type planets. Finally, we conclude the paper and give outlooks for the future in Sec. \ref{sec:summary}.

\section{Hypotheses and measurement setting}
\label{sec:preliminaries}
\begin{figure}
    \centering
    \includegraphics[width=1\linewidth]{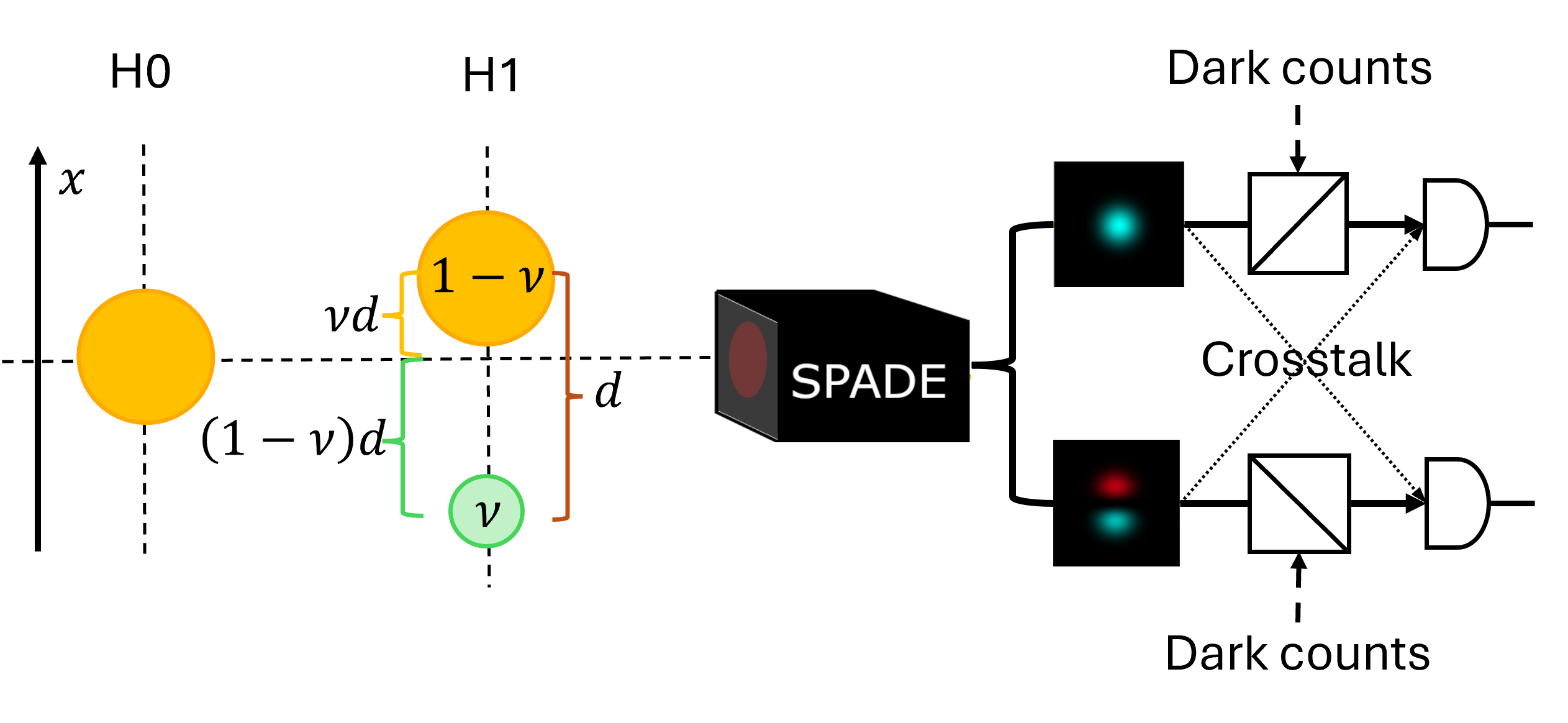}
    \caption{Schematic representation of the considered measurement scenario. Under hypothesis H0, there is only one source. Under hypothesis H1, there are two sources separated by distance $d$ and characterized by relative brightness $\nu$, with the same total brightness as the source from hypothesis H0. In both hypotheses, the coordinate system is aligned with the system's center of brightness. The SPADE measurement is vulnerable to two forms of experimental noise: crosstalk, i.e., the possibility of measuring a photon in the incorrect mode, and dark counts, i.e., the possibility of measuring a photon not originating from the studied system.}
    \label{fig:scheme}
\end{figure}

We begin with a detailed description of the two hypotheses (see Fig. \ref{fig:scheme}): According to hypothesis H1, two weak, incoherent light sources (e.g. faraway thermal sources) of relative brightnesses $\nu$ and $1-\nu$ are separated by a distance $d$ in the object plane. Here, $\nu\in [0,1]$, with $\nu=1/2$ corresponding to equal brightnesses and $\nu=0,1$ corresponding to only one visible source. The coordinate system is chosen to be such that the two sources are aligned with its $x$-axis, with the position of the source of brightness $\nu$ given by $x_{\textnormal{ex}}=-(1-\nu) d$ and the other source by $x_{\textnormal{st}}=\nu d$. In this way, the origin of the coordinate system lies precisely in the center of brightness of the system:
\begin{equation} \label{eq:center_of_brightness}
    0 = \nu \,x_{\textnormal{ex}} + (1-\nu) x_{\textnormal{st}},
\end{equation}
which can be efficiently predetermined using intensity measurements. We remark that throughout the rest of the manuscript, we will presume $\nu \ll 1$, so that the source at $x_{\textnormal{ex}}$ can be identified with the exoplanet, and the source at $x_{\textnormal{st}}$ with its home star, hence the notation.

We consider a diffraction-limited imaging system with a Gaussian point spread function of width $w$:
\begin{equation} \label{eq:PSF}
    u_{0}(x)=\sqrt[4]{\frac{2}{\pi w^2}}e^{-x^2/w^2},
\end{equation}
so that the spatial distribution of the electromagnetic field in the image plane coming from a source at $x_0$ is given by $u_{0}(x-x_0)$ \cite{goodman1985}. For weak sources, most of the photon detection events are single-photon events. Therefore, under hypothesis H1, the results in the image plane are effectively described by repeated measurements on $N$ copies of the single-photon state \cite{superresolution_Tsang_2016}  
\begin{equation} \label{eq:state_H1}
    \hat{\rho}_{\textnormal{H1}}(\nu,d)\approx 
        \nu \ket{\phi(x_{\textnormal{ex}})}\bra{\phi(x_{\textnormal{ex}})}
        + (1-\nu) \ket{\phi(x_{\textnormal{st}})}\bra{\phi(x_{\textnormal{st}})},
\end{equation} where  
$\ket{\phi(x_0)}=\int dx\;u_{0}(x-x_0)\ket{x}$
and $\ket{x}$ stands for the single-photon position eigenstate in the image plane. 

According to hypothesis H0, there is only one source in the object plane, characterized by the same total brightness as the two sources from hypothesis H1, and located in their center of brightness (i.e., at $x=0$). In this case, the measurement results are effectively described by repeated measurements on the state
\begin{equation} \label{eq:state_H0}
      \hat{\rho}_{\textnormal{H0}} = \lim_{d\to 0}\hat{\rho}_{\textnormal{H1}}(\nu,d)
        =\ket{\phi(0)}\bra{\phi(0)},
\end{equation}
where we note that the limit $d\to 0$ necessarily implies lack of dependence on $\nu$. We remark that throughout the paper, we will usually omit writing the explicit formulas for quantities associated with hypothesis H0, since, just like the density operator above, they are always defined by taking the limit $d\to 0$ for hypothesis H1.

\section{Optimal source discrimination}
\label{sec:testing_optimal}
Any decision strategy involving two hypotheses is susceptible to two types of errors. With so-called \emph{probability of error of the first kind} $\alpha$, i.e., the \emph{false alarm probability}, hypothesis H1 can be assumed when hypothesis H0 is true. Conversely, with \emph{probability of error of the second kind} $\beta$, i.e., the \emph{miss probability}, hypothesis H0 can be assumed when hypothesis H1 is true. In \emph{asymmetric hypothesis testing}, the efficiency of the test is determined solely by the miss probability, which is to be minimized even at the cost of the false alarm probability, provided the latter does not exceed some set threshold value $\bar{\alpha}$ \cite{hypothesis_testing_book_Van_Trees_2001}. This is the right approach whenever the two types of error are not equally important, e.g., in military scenarios, where failing to detect an enemy attack is more costly than a false alarm \cite{hypothesis_testing_Sorelli_2021}. Similarly, in the case at hand, because exoplanets are rare, it is preferable to occasionally falsely detect one than to miss it \cite{Exo_4}.

We now proceed to summarize the state-of-the-art regarding optimal source discrimination in the considered context. To a good approximation, in the asymptotic limit of large sample sizes $N$, the optimal miss probability characterizing any given measurement strategy decays exponentially as \cite{quantum_Stein_lemma_Hiai_1991,quantum_Stein_lemma_Ogawa_2000,asymmetric_miss_probability_Tomamichel_2013,asymmetric_miss_probability_Li_2014,Exo_4}
\begin{align} \label{eq:beta_asymptotic_decay}
    \beta(N) \approx 
        \exp\left[-N D(\textnormal{H0}|\textnormal{H1})\right],
\end{align}
where
\begin{align} \label{eq:D_general}
    D({\textnormal{H0}}|{\textnormal{H1}}) 
        \coloneqq \sum_k p(k|\textnormal{H0})
        \ln \frac{p(k|\textnormal{H0})}{p(k|\textnormal{H1})}
\end{align}
is the (classical) \emph{relative entropy} \cite{relative_entropy_original_Kullback_1951,information_theory_review_Witten_2020}, a common measure of statistical distinguishability between two probability distributions. In the case at hand, $p(k|\textnormal{H})$ denote the probability of obtaining measurement outcome $k$ under hypothesis H, and so the relative entropy quantifies the distinguishability between the two hypotheses. 

By considering the quantum generalization of Eq. (\ref{eq:beta_asymptotic_decay}), i.e., with the relative entropy replaced by the quantum relative entropy between the two states (\ref{eq:state_H1}-\ref{eq:state_H0}), the best possible scaling of the miss probability allowed by quantum mechanics was derived in Ref. \cite{Exo_4}. Assuming small relative brightnesses $\nu\ll 1$ and separations deep within the \emph{sub-Rayleigh} regime 
\begin{align} 
    s \coloneqq d/w\ll 1 
\end{align}
the quantum relative entropy was shown to equal \cite{Exo_4}
\begin{align} \label{eq:D_Q}
    D_{\textnormal{Q}} \approx \nu s^2.
\end{align}
Operationally, the quantum relative entropy can be viewed as the classical relative entropy (\ref{eq:D_general}) maximized over all the possible measurements. In the case at hand, it was proven to be attained by \emph{spatial demultiplexing (SPADE)} followed by intensity measurements in the Hermite-Gauss modes \cite{lasers_1986_siegman}:
\begin{align} \label{eq:HG_modes}
    u_{n}(x) \coloneqq 
        \frac{H_n(\sqrt{2}x/w)}{\sqrt{2^{n}n!}}u_{0}(x),
\end{align}
where $H_n(z)$ are the Hermite polynomials and $u_{0}(x)$ is  the Gaussian point spread function (\ref{eq:PSF}). Under ideal conditions, the probability of measurement outcome $n$, i.e., detecting a photon in mode $u_{n}$, equals \cite{unbalanced_sources_crosstalk_Linowski_2023,hypothesis_testing_Schlichtholz_2024}
\begin{align} \label{eq:p_SPADE}
\begin{split}
    p_{\textnormal{SD}}(n|\textnormal{H1}) = 
        \nu|\gamma_{n}(x_{\textnormal{ex}})|^2
        + (1-\nu)|\gamma_{n}(x_{\textnormal{st}})|^2.
\end{split}
\end{align}
Here, $\gamma_{n}$ denote the mode overlaps,
\begin{align} \label{eq:f_nm}
    \gamma_{n}(x_0) \coloneqq \int_{-\infty}^\infty dx \,u_n(x-x_0)\,u_0(x) =
        \frac{(x_0/w)^n}{\sqrt{n!}}e^{-x_0^2/2w^2},
\end{align}
and we remark that in equations, we denote SPADE-related quantities by the shorter SD for better readability. Substituting the above quantities into the formula (\ref{eq:D_general}) for the relative entropy, and expanding to the lowest non-zero order in $\nu$ and $s$, one obtains \cite{Exo_4}
\begin{align} 
    D_{\textnormal{SD}} \approx D_{\textnormal{Q}}.
\end{align}

To put this value into perspective, it was compared to its analog obtained for conventional imaging, based on spatially resolved intensity measurements \cite{Exo_4}. Assuming such \emph{direct imaging} to be ideal, i.e., noiseless and continuous, the measurement outcome $x$ simply denotes the position of the measured photon on the screen, with the associated continuous probability distribution
\begin{align} \label{eq:p1_DI}
\begin{split}
    p_{\textnormal{DI}}(x|{\textnormal{H1}}) 
        &\coloneqq \bra{x}\hat{\rho}_{\textnormal{H1}}(\nu,d)\ket{x} \\
        &= \nu\,u_{0}^2(x-x_{\textnormal{ex}})
        + (1-\nu)\,u_{0}^2(x-x_{\textnormal{st}}).
\end{split}
\end{align}
Consequently, summation in Eq. (\ref{eq:D_general}) for the relative entropy is promoted to integration:
\begin{align} \label{eq:D_DI_definition}
    D_{\textnormal{DI}}
        = \int_{-\infty}^{\infty} dx \, p_{\textnormal{DI}}(x|\textnormal{H0})
        \ln \frac{p_{\textnormal{DI}}(x|\textnormal{H0})} 
        {p_{\textnormal{DI}}(x|\textnormal{H1})}.
\end{align}
Expanding the integrand to the lowest non-zero order in $\nu$ and $s$ and integrating, we obtain
\begin{align} \label{eq:D_DI}
    D_{\textnormal{DI}}
        \approx 4 \nu^2 s^4,
\end{align}
which is smaller than the optimal value (\ref{eq:D_Q}) by one order in $\nu$ and two in $s$, demonstrating a clear disadvantage in comparison to SPADE \footnote {We remark that in Ref. \cite{Exo_4}, a different value was obtained, $D_{\textnormal{DI}} \approx \nu^2 s^2$, which arised from centering the coordinate system for hypothesis H1 on the star, rather than the center of brightness. Note that translational invariance of direct imaging is preserved in both cases.}. 


\section{Impact of noise}
\label{sec:testing_practical}
So far, we have assumed all the measurements to be ideal, which, as discussed in the Introduction, is never the case in reality. To assess the performance of SPADE under realistic conditions, we consider two forms of experimental imperfections. Firstly, due to \emph{crosstalk} \cite{metrology_noise_Nichols_2016,noise_Oh_2021,Sorelli_practical_superresolution_2021,superresolution_no_location_Grace_2020,Almeida_misalignment_2021}, the actual measurement basis deviates from the ideal Hermite-Gauss basis as \cite{crosstalk_original_PRL,unbalanced_sources_crosstalk_Linowski_2023,hypothesis_testing_Schlichtholz_2024}:
\begin{align} \label{eq:crosstalk_definition}
    u_{n}(x) \to \sum_{k=0}^{D-1} C_{nk}u_{k}(x).
\end{align}
Here, $C$ stands for a (unitary) crosstalk matrix, which is close to unity, and $D$ restricts the number of measured modes, since in reality we always measure a finite number of them \cite{crosstalk_original_PRL}. Because of crosstalk, the mode overlaps (\ref{eq:f_nm}) are modified as
\begin{align} \label{eq:f_nm_crosstalk} 
    \gamma_{n}(x) \to \sum_{k=0}^{D-1} C_{nk} \gamma_{k}(x).
\end{align}
Irrespectively of the presence of crosstalk, the restriction to $D$ modes necessitates renormalization of the probabilities:
\begin{align} \label{eq:p_SPADE_renormalization} 
    p_{\textnormal{SD}}(n|\textnormal{H1}) \to \frac{p_{\textnormal{SD}}(n|\textnormal{H1})}{\sum_{k=0}^{D-1}p_{\textnormal{SD}}(k|\textnormal{H1})}.
\end{align}

Secondly, due to \emph{dark counts} \cite{superresolution_limits_SPADE_Len_2020,imaging_noisy_Lupo_2020}, the detector will occasionally measure photons that do not originate from the probed system, mostly ones generated by the detector's internal thermal noise. Expressing the probabilities (\ref{eq:p_SPADE}) explicitly in terms of the number $N_{n}$ of photons measured in mode $u_{n}$, the effect of dark counts amounts to the following modification:
\begin{align} \label{eq:N_nm_dark_counts} 
\begin{split}
    p_{\textnormal{SD}}(n|\textnormal{H1}) = \frac{N_{n}}{N}\to \frac{N_{n} 
        + N_{\textnormal{dark}}}{N + D N_{\textnormal{dark}}}
\end{split}
\end{align}
and analogously for hypothesis H0. For the clarity of presentation, we assumed that each detector mode experiences approximately the same number $N_{\textnormal{dark}}$ of dark counts (this assumption can be lifted with relative ease). 

Taking both forms of noise into account, we substitute the probabilities for noisy SPADE into Eq. (\ref{eq:D_general}) and expand to the lowest non-zero order in $\nu,s\ll1$, assuming that the total amount of noise in the system is much smaller than the signal. Proceeding in this way, we obtain the following formula for the relative entropy:
\begin{align} \label{eq:D_SPADE_general}
    D_{\textnormal{SD}} \approx
        p_s - p_0\left(1 + \ln \frac{p_s}{p_0}\right),
\end{align}
The expression depends solely on the probability of measuring a photon in mode $u_{1}$, approximately equal to (truncated to leading orders in all effects)
\begin{align} \label{eq:p_10_noisy} 
\begin{split}
    p_s\coloneqq p_{\textnormal{SD}}(n|\textnormal{H1}) 
        \approx p_{\textnormal{cross}} + p_{\textnormal{dark}} + \nu s^2,
\end{split}
\end{align}
where $p_\textnormal{cross}\coloneqq|C_{10}|^2$ is the probability of crosstalk from mode $u_{0}$ to mode $u_{1}$ and 
\begin{align} \label{eq:p_dark} 
    p_{\textnormal{dark}} \coloneqq 
        \frac{N_{\textnormal{dark}}/N}{1+D N_{\textnormal{dark}}/N}
\end{align}
is the mean probability of dark counts per mode. Note that in the newly introduced shorthand notation,
\begin{align}
     p_0 \coloneqq p_{\textnormal{SD}}(n|\textnormal{H0}) = \lim_{s\to 0}p_s \approx p_{\textnormal{cross}} + p_{\textnormal{dark}},
\end{align}
which conveniently quantifies the total amount of noise in the system.

The explicit form of scaling of the efficiency of SPADE depends on the relation between the amount of noise in the measurement apparatus and the parameters of the probed system:
\begin{align} \label{eq:D_SPADE_regimes}
    D_{\textnormal{SD}} \approx \begin{dcases}
        \nu s^2 & \nu s^2 \gg p_0,\\
        \frac{\nu^2 s^4}{2p_0} & \nu s^2 \ll p_0,
    \end{dcases}
\end{align}
where the two cases follow from Eq. (\ref{eq:D_SPADE_general}) depending on whether we expand first in $p_0$ (top line) or $\nu$ and $s$ (bottom line). As seen, the results are approximately ideal (\ref{eq:D_Q}) as long the amount of noise is relatively small, and far worse than ideal if the amount of noise is not negligible. 

This dual behavior mirrors analogous results for separation estimation \cite{unbalanced_sources_crosstalk_Linowski_2023} and symmetric hypothesis testing \cite{hypothesis_testing_Schlichtholz_2024}, but with one key difference: Due to the dependence of the scaling criterion on the relative brightness, here, the nearly ideal regime provided in the top line is virtually unachievable in practice. Indeed, if we assume the separation to be as large as $s=0.1$, and take the largest relative brightness reported in the context of exoplanet detection in Ref. \cite{coronagraphs_perfect_Deshler_experiment_2024}, $\nu = 10^{-5}$, the nearly ideal regime would still require the combined probability of crosstalk and dark counts to fall below $p_0 = 10^{-7}$, multiple orders of magnitude below values currently reported experimentally \cite{Boucher:20}. This means that at least in the near term, the practical efficiency of SPADE for exoplanet detection in the sub-Rayleigh regime is unambiguously captured by the bottom line of Eq. (\ref{eq:D_SPADE_regimes}).

\begin{figure}[!t]
    \centering
    \includegraphics[width=.48\textwidth]{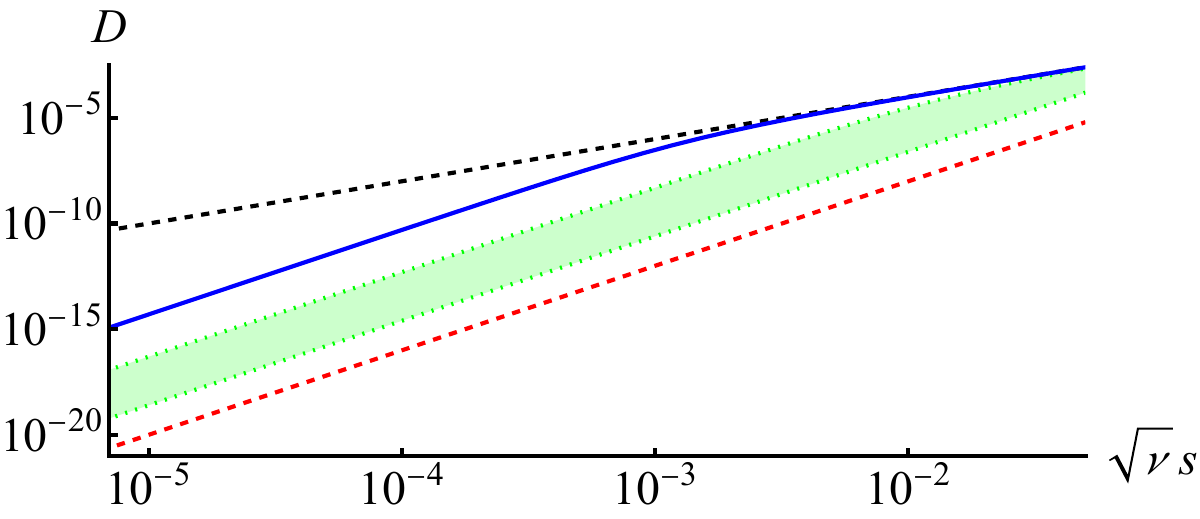}
    \caption{Comparison of relative entropies $D$ as a function of relative brightness-rescaled separation $\sqrt{\nu}s$. The top black and bottom red dashed curves correspond to the quantum bound (\ref{eq:D_Q}) and ideal direct imaging (\ref{eq:D_DI}), respectively. The green band corresponds to noisy SPADE (\ref{eq:D_SPADE_general}) with noise strength varying from low, $p_0=10^{-4}$, up to large, $p_0=0.02$. 
    For realistic noise strengths, the scaling of SPADE is the same as for direct imaging, albeit with a much larger scaling coefficient, here resulting in from one up to four orders of magnitude of advantage in the value of the relative entropy. Through Eq. (\ref{eq:beta_asymptotic_decay}), this translates to several orders of magnitude of advantage in the required sample size $N$ for a fixed target miss probability. For reference, the blue solid line shows how SPADE scales for the very low noise strength $p_0=10^{-6}$, illustrating the dual behavior~(\ref{eq:D_SPADE_regimes}).}
    \label{fig:D}  
\end{figure}

Still, even noisy SPADE performs markedly better than ideal direct imaging (\ref{eq:D_DI}) -- although both are characterized by the same scaling powers, the scaling coefficient for SPADE is far superior for experimentally expected noise strengths: $1/2p_0 \gg 4$. Moreover, since this coefficient is monotonically increasing with decreasing $p_0$, even if the near-ideal scaling is not achievable for SPADE, it can be gradually approached through noise mitigation. This is illustrated in Fig. \ref{fig:D}, a comparison of the relative entropy for noisy SPADE with the quantum bound and ideal direct imaging, where by minimizing noise, SPADE can achieve several orders of magnitude of advantage over direct imaging.


\section{Comparison with coronagraphs}
\label{sec:coronagraphs}
As discussed previously, conventional imaging employed in contemporary astronomy is more refined than standard direct imaging. A particularly prominent family of measurements are \emph{coronagraphs} \cite{coronagraphs_Soummer_2003,coronagraphs_Guyon_2006,coronagraphs_review_2018,coronagraphs_NASA_1_Kasdin_2023}, which rely on a variety of techniques, such as signal masking and apodization, to increase the effective relative visibility of the exoplanet. Although the details of realization depend on the specific method, see, e.g., the classic Lyot \cite{Lyot_coronagraph_Soummer_2007}, vortex \cite{coronagraphs_vortex_Foo_2005} or PIAACMC \cite{coronagraphs_PIAACMC_Guyon_2010} coronagraphs, the end effect boils down to minimizing the detection of the fundamental mode of the system, which is predominantly occupied by the starlight. Following \cite{coronagraphs_perfect_Deshler_theory_2024}, we define the coronagraph operator as
\begin{align}
        \hat{\mathcal{C}} \coloneqq \hat{\mathds{1}} - R\ket{\phi(0)}\bra{\phi(0)},
\end{align}
where $\braket{x|\phi(0)}=u_0(x)$ [see the definition below Eq. (\ref{eq:state_H1})] and we introduced $R\in[0,1)$ to denote the portion of the fundamental mode successfully removed from the measurement.

The coronagraphic probability distributions are then similar as for direct imaging (\ref{eq:p1_DI}), except the state is projected using $\hat{\mathcal{C}}$, i.e.,
\begin{align} \label{eq:p1_CG}
\begin{split}
    p_{\textnormal{CG}}(x|{\textnormal{H1}}) 
        \coloneqq \bra{x}\hat{\mathcal{C}}\hat{\rho}_{\textnormal{H1}}(\nu,d)\hat{\mathcal{C}}^\dag\ket{x}.
\end{split}
\end{align}
One key difference in comparison to direct imaging is that, since $\hat{\mathcal{C}}$ is not unitary, the above distribution is not normalized. This prompts us to consider the probability that a photon is not detected by the coronagraph:
\begin{align} \label{eq:p1_CG}
\begin{split}
    p_{\textnormal{CG}}'({\textnormal{H1}}) 
        = 1 - \int_{-\infty}^{\infty} dx \,p_{\textnormal{CG}}(x|{\textnormal{H1}}),
\end{split}
\end{align}
which serves purely as a ``bookkeeping'' tool, so that the corresponding relative entropy is well-defined:
\begin{align} \label{eq:D_coronagraphs_definition}
\begin{split}
    D_{\textnormal{CG}} = \int_{-\infty}^{\infty} 
        dx \, p_{\textnormal{CG}}(x|&\textnormal{H0})
        \ln \frac{p_{\textnormal{CG}}(x|\textnormal{H0})} 
        {p_{\textnormal{CG}}(x|\textnormal{H1})}\\
        &+ p_{\textnormal{CG}}'(\textnormal{H0})
        \ln \frac{p_{\textnormal{CG}}'(\textnormal{H0})} 
        {p_{\textnormal{CG}}'(\textnormal{H1})}.
\end{split}
\end{align}
We stress that the considered model of coronagraphs is highly idealized: the only imperfection it assumes is that the fundamental mode is not removed perfectly, $R\neq 1$, and it does not take into account other forms of noise.

Calculating the above explicitly, we find that, upon expansion in $\nu$, $s$ and $1-R$ up to first non-zero order, the idealized coronagraph turns out to be qualitatively similar to noisy SPADE (\ref{eq:D_SPADE_regimes}):
\begin{align} \label{eq:D_coronagraphs_R}
    D_{\textnormal{CG}} \approx \begin{dcases}
        \nu s^2 & \nu \gg 1-R,\\
        \frac{3\nu^2 s^4}{2(1-R)^2} & \nu \ll 1-R,
    \end{dcases}
\end{align}
where the two cases depend on whether we expand first in $1-R$ (top line) or $\nu$ (bottom line). On the one hand, in the top line, we observe the quantum-optimal efficiency (\ref{eq:D_Q}) that one should theoretically expect from the so-called \emph{perfect coronagraph}, which cuts out nearly all the photons from the fundamental mode \cite{coronagraphs_perfect_Deshler_theory_2024,coronagraphs_perfect_Deshler_experiment_2024}. On the other hand, in the bottom line, corresponding to a more realistic coronagraph, we see the same inferior scaling in relative brightness and separation as for realistic SPADE, similarly improved by a division by a very small constant.

Having no access to the expected set of values of the abstract parameter $R$ (knowing only that it should be close to one), we cannot compare the relative entropy of the idealized coronagraph to that of SPADE quantitatively. To do so, we need to consider the impact of noise on coronagraphs too. In this regard, it will be sufficient to restrict ourselves to dark counts. Under their influence, the coronagraphic distribution (\ref{eq:p1_CG}) changes as
\begin{align} \label{eq:p1_CG_dc}
\begin{split}
    p_{\textnormal{CG}}(x|{\textnormal{H1}}) \to
        (1-\mu)p_{\textnormal{CG}}(x|{\textnormal{H1}})
        + \mu f_{\textnormal{dark}}(x),
\end{split}
\end{align}
where $\mu\ll 1$ quantifies the relative total probability of detected dark counts, while $f_{\textnormal{dark}}(x)$ is the dark count probability function. Similarly as previously for SPADE, we assume that the probability of dark counts across the whole spectrum of detected photons is uniform, and so their probability is naturally modeled by the flat distribution:
\begin{align} \label{eq:f_dark_counts}
\begin{split}
    f_{\textnormal{dark}}(x) \coloneqq \begin{dcases}
        1/L & x\in[-L/2,L/2],\\
        0 & \textnormal{otherwise}.
    \end{dcases}
\end{split}
\end{align}
Here, the parameter $L$ should be understood as the effective size of the camera, and thus to be large enough to cover the part of the image plane in which the Gaussian distributions associated with the sources are significantly different from zero. Given $L$, it is convenient to recast the total probability of dark counts as 
\begin{align} \label{eq:mu_decomposition}
    \mu=\varrho L,
\end{align}
with $\varrho$ denoting the (approximately $L$-independent) probability of dark counts per unit length (in units of point spread function width $w$) \footnote{Note that, in general, in analogy to Eq. (\ref{eq:p_dark}) for SPADE, $\mu = n L/(N+n L)$, where $n$ is the density of dark counts. This means that, strictly speaking, $\varrho$ depends on $L$, however, since $\mu\ll 1$, $\varrho = n/(N+n L) \approx n/N$, which is $L$-independent.}.


Recalculating the coronagraphic relative entropy (\ref{eq:D_coronagraphs_definition}) with dark counts included, we find that, in the leading order in $\nu$, $s$, $1-R$, and $\varrho$, and for asymptotically large~$L$,
\begin{align} \label{eq:D_coronagraphs_dark_counts}
    D_{\textnormal{CG,dark}} = \frac{3\nu^2s^4}{8\sqrt{\pi}\varrho}.
\end{align}
Comparing with noisy SPADE (\ref{eq:D_SPADE_regimes}), we find that the latter is superior to coronagraphs if 
\begin{align} \label{eq:SPADE_superiority_condition}
    p_0 = p_{\textnormal{cross}} + p_{\textnormal{dark}} < \frac{4\sqrt{\pi}}{3}\varrho\approx 2.36 \varrho.
\end{align}
To see why this condition should be fulfilled in reality, we first observe that $p_{\textnormal{dark}}$ is the probability of dark counts in SPADE per mode, i.e., per one single-photon detector. Since the sources are contained within an interval of the order of unit length, the image captured by coronagraphic camera will not be able to efficiently distinguish between the hypotheses unless the number $m$ of detectors (pixels) per unit length is much greater than one. For simplicity, let us assume that all of the coronagraphic detectors are of the same quality as the ones used for SPADE. In reality, high-quality detectors, e.g., superconducting nanowire detectors \cite{SPD}, should be more affordable and technically feasible for SPADE, as it requires far fewer of them, and does not need to pack them as densely. Then, $\varrho \approx m p_{\textnormal{dark}}$, and so Eq. (\ref{eq:SPADE_superiority_condition}) becomes
\begin{align} \label{eq:SPADE_superiority_condition_refined}
    p_{\textnormal{cross}}\lessapprox \left(2.36\,m-1\right)p_{\textnormal{dark}},
\end{align}
which is easily fulfilled for any realistic $m$ and noise strengths. It is worth adding that regardless of $m$, there always exists a regime (e.g., for sufficiently faraway and thus faint sources) in which the signal to noise ratio is so low that the probability of dark counts dominates over crosstalk.


In any case, the condition (\ref{eq:SPADE_superiority_condition_refined}) is likely too strict, with the efficiency of real coronagraphs further reduced by additional factors, such as finitely sized pixels. In this regard, it is instructive to go back to the perfect coronagraph, the only known quantum-optimal coronagraph \cite{coronagraphs_perfect_Deshler_experiment_2024}. By construction, the experimental realization of the perfect coronagraph is equivalent to a standard SPADE scheme followed by additional processing (e.g., mode multiplexing) required to convert the result to a conventional image \cite{coronagraphs_perfect_Deshler_theory_2024,coronagraphs_perfect_Deshler_experiment_2024}. From the point of view of measurement efficiency, the latter only contributes to increasing the measurement noise, and so the perfect coronagraph necessarily performs worse than the standard SPADE measurement.

\section{Decision strategy}
\label{sec:test_explicit}
The analysis of the probability of error for SPADE remains only a theoretical curiosity if we do not know the associated decision strategy. Formally speaking, the optimal decision strategy for any measurement scheme, i.e., the one corresponding to the exponential decay (\ref{eq:beta_asymptotic_decay}), is given by the \emph{likelihood ratio test}, which adapted to SPADE, reads as
\begin{align} \label{eq:likelihood_ratio_symmetric} 
    \frac{P_N\left(\{N_{n}\}|\textnormal{H1}\right)}
    {P_N\left(\{N_{n}\}|\textnormal{H0}\right)} 
        \underset{H0}{\overset{H1}{\gtrless}} 
        \lambda_N\left(\{N_{n}\},\bar{\alpha}\right).
\end{align}
Here, $P_N\left(\{N_{n}\}|\textnormal{H}\right)$ is the probability of obtaining a given set of photon counts $\{N_{n}\}$ given hypothesis H, and $\lambda_N$ is a threshold function of the assumed maximum tolerated false detection probability $\bar{\alpha}$. The symbol $\underset{H0}{\overset{H1}{\gtrless}}$ indicates that the assumed hypothesis should be H1 if the l.h.s. is strictly greater than the r.h.s., and H0 otherwise. 

Unfortunately, the likelihood ratio test has significant drawbacks limiting its practical applicability \cite{hypothesis_testing_Schlichtholz_2024}. Most significantly, $P_N\left(\{N_{n}\}|\textnormal{H1}\right)$ inevitably depends on both the separation and relative brightness, meaning that to perform the test, the latter parameters have to be first estimated from the data, using, e.g., the method of moments \cite{method_of_moments_Sorelli_2021}. Errors in this estimation deviate this decision strategy from optimality, and can lead to overestimating its efficiency. Moreover, there are very few cases in which the threshold $\lambda_N$ can be expressed in a closed form \cite{hypothesis_testing_Sorelli_2021}, reducing the intuitive understanding of the decision strategy.

For these reasons, we follow the recent efforts in symmetric hypothesis testing \cite{naive_test_Shapiro_2018,hypothesis_testing_Schlichtholz_2024} and simplify the problem by restricting ourselves to only two single-measurement outcomes: (1) photon in the Hermite-Gauss mode $u_{1}$, and (2) photon in any other mode. This is justified by the fact that for small separations, almost all of the information about the difference between the hypotheses is contained in mode $u_{1}$ \cite{naive_test_Shapiro_2018,hypothesis_testing_Schlichtholz_2024}, as evidenced in the case at hand by Eq.~(\ref{eq:D_SPADE_general}), where only the noise in this mode appears. In this binary SPADE scenario, the likelihood ratio test (\ref{eq:likelihood_ratio_symmetric}) is equivalent to a single inequality for $N_1$:
\begin{align} \label{eq:N10_generic_test}
    N_{1} \underset{H0}{\overset{H1}{\gtrless}} \bar{N}_{1}
        \left(N,\nu,s,p_0\right),
\end{align}
since for a fixed number of samples, the number of photons $N_{1}$ in mode $u_{1}$ remains the only data gathered in the experiment. As denoted above, the threshold $\bar{N}_{1}$ can in principle depend on the number of samples, relative brightness, separation, as well as detector noise.

Since the scheme permits only two single-photon outcomes, the expressions for the error probabilities implied by the test (\ref{eq:N10_generic_test}) are completely analogous to those for flipping an unfair coin $N$ times, and follow the binomial distribution:
\begin{align} 
    \alpha(N) &= \sum_{\bar{N}_{1}<k\leqslant N}\binom{N}{k} p_0^k\left(1-p_0\right)^{N-k}, 
    \label{eq:alpha_general}\\
    \beta(N) &= \sum_{0\leqslant k \leqslant \bar{N}_{1}}^{}\binom{N}{k} p_s^k\left(1-p_s\right)^{N-k}. 
    \label{eq:beta_general}
\end{align}
Here, $\alpha(N)$ is simply the probability of obtaining more photons in mode $u_{1}$ than expected under the threshold $\bar{N}_{1}$ for hypothesis H0, and similarly, $\beta(N)$ is the probability of obtaining fewer photons in mode $u_{1}$ than expected for hypothesis H1.

Before we consider the noisy scenario, let us briefly discuss the result for ideal SPADE. In this case, $p_0=0$ for any $\bar{N}_{1}$, implying $\alpha(N)=0$ and therefore trivializing the basic idea behind asymmetric hypothesis testing, since in this case it makes no sense to put any upper bound on the false alarm probability. The miss probability $\beta(N)$ is then minimized by the trivial threshold $\bar{N}_{1}=0$, and the test boils down to assuming hypothesis H1 whenever even one photon is measured in mode $u_1$, just like for ideal symmetric hypothesis testing~\cite{naive_test_Shapiro_2018}.

In the presence of noise, this changes drastically. As discussed previously, in asymmetric hypothesis testing, one demands that the value of $\alpha(N)$ does not exceed some chosen maximum $\bar{\alpha}$, and then looks for a decision strategy minimizing $\beta(N)$ under this condition. Because the decision strategy at hand depends on only one free parameter $\bar{N}_{1}$, demanding $\alpha(N)\leqslant\bar{\alpha}$ in Eq. (\ref{eq:alpha_general}) fixes this parameter, determining the strategy completely. 

The calculation can be performed analytically if we assume the standard Gaussian approximation to the binomial distribution \cite{statistics_book_Bertsekas_2008}
\begin{align} \label{eq:Gaussian_approximation}
\begin{split}
    \binom{N}{k} p_s^k\left(1-p_s\right)^{N-k} \approx 
        \frac{1}{\sqrt{\pi\sigma_s^2}}e^{-\left(k-Np_s\right)^2/\sigma_s^2},
\end{split}
\end{align}
which is almost exact for large sample sizes due to the approximation error scaling as $1/\sqrt{N}$ \cite{statistics_book_Wasserman_2004}. Using the above, we find that the threshold determining the test (\ref{eq:N10_generic_test}) equals
\begin{align} \label{eq:N_threshold}
\begin{split}
    \bar{N}_{1} = \: & N p_0 + \sigma_0\erf^{-1}\left(1-2\bar{\alpha}\right),
\end{split}
\end{align}
where $\erf$ is the error function, $\erf^{-1}$ is its inverse, and $\sigma_s^2\coloneqq 2N p_s(1-p_s)$. We observe that because the false alarm probability $\alpha(N)$ is calculated assuming hypothesis H0, the obtained threshold does not depend on neither the relative brightness nor the separation, constituting an advantage over the likelihood ratio test (\ref{eq:likelihood_ratio_symmetric}).

The associated miss probability follows from Eq. (\ref{eq:beta_general}). Explicitly (again, under the Gaussian approximation), we have:
\begin{align} \label{eq:beta_explicit}
\begin{split}
    \beta(N) = \frac{1}{2}\left[\erf\left(\frac{N p_s}{\sigma_s}\right)
        + \erf\left(
        \frac{\bar{N}_{1}-N p_s}{\sigma_s}\right)\right].
\end{split}
\end{align}
Note that, unlike the test, the miss probability does depend on both the relative brightness $\nu$ and the separation $s$ (through $p_s$). However, by construction, the probability of error is necessarily increasing as $\nu$ and $s$ decrease. Therefore, even in the absence of precise knowledge about them, the miss probability can be meaningfully upper-bounded by using Eq. (\ref{eq:beta_explicit}) with the smallest $\nu$ and $s$ expected for the system of interest (e.g., for Earth-like exoplanets, $\nu \gtrapprox 10^{-11}$ \cite{coronagraphs_review_2018,coronagraphs_perfect_Deshler_experiment_2024}). 

\begin{figure}[!t]
    \centering
    \begin{subfigure}[b]{0.49\textwidth}
        \centering
        \includegraphics[width=\textwidth]{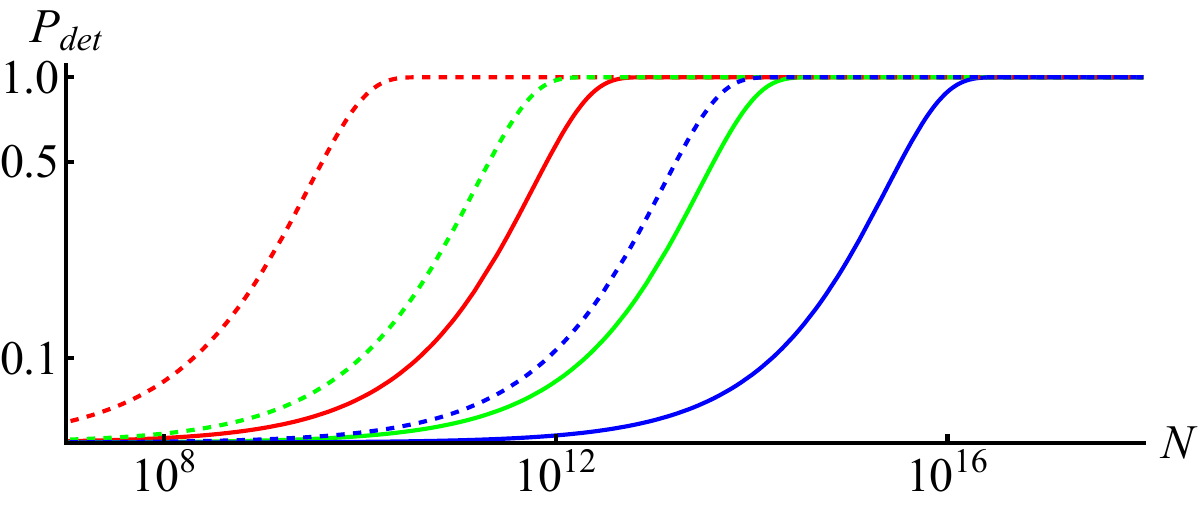}
    \end{subfigure}\\
    \begin{subfigure}[b]{0.49\textwidth}
        \centering
        \includegraphics[width=\textwidth]{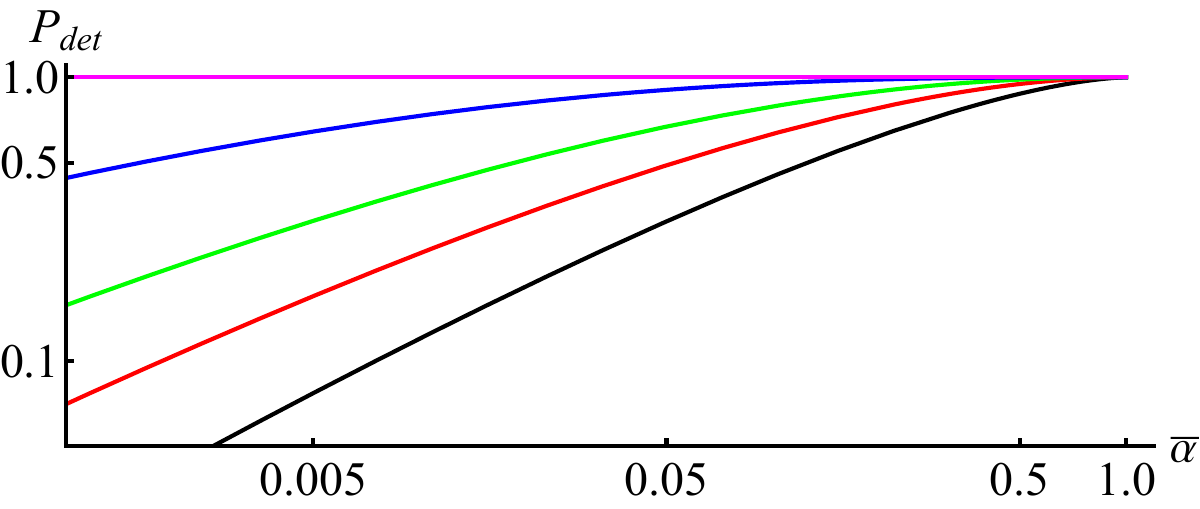}
    \end{subfigure}
    \caption{Top: Detection probability $P_{\textnormal{det}}$ as a function of sample size $N$ for the typical threshold false alarm probability $\bar{\alpha}=0.05$ and relative brightness $\nu=10^{-5}$ (corresponding to Hot Jupiters \cite{coronagraphs_perfect_Deshler_experiment_2024}). Solid (dashed) curves were drawn assuming large noise strength, $p_0=0.02$ (small noise strength, $p_0=10^{-4}$) \cite{unbalanced_sources_crosstalk_Linowski_2023,hypothesis_testing_Schlichtholz_2024} and correspond to different separations: $s=\{0.16,0.06,0.02\}$ (left to right) \cite{crosstalk_original_PRL,Boucher:20,unbalanced_sources_crosstalk_Linowski_2023,hypothesis_testing_Schlichtholz_2024}. Bottom: Detection probability $P_{\textnormal{det}}$ as a function of threshold false alarm probability $\bar{\alpha}$ for sample size $N=2\cdot 10^{13}$ (the same order of magnitude as in the proof of principle experiment \cite{SPADE_superresolution_experiment_Rouviere_2024}), relative brightness $\nu=10^{-5}$ and separation $s=0.06$. From top to bottom, the curves correspond to noise strengths $p_0=\{0.0001,0.003,0.006,0.01,0.02\}$.}
    \label{fig:P_det}  
\end{figure}

To assess the practical efficiency of the derived test, in Fig. \ref{fig:P_det} we plot the \emph{detection probability}
\begin{align} \label{eq:P_det_explicit}
\begin{split}
    P_\textnormal{det}(N) \coloneqq 1 - \beta(N)
\end{split}
\end{align}
as a function of sample number $N$ and as a function of threshold false alarm probability $\bar{\alpha}$ for a variety of experimentally motivated values of system parameters and noise strengths \cite{SPADE_superresolution_experiment_Rouviere_2024,crosstalk_original_PRL,unbalanced_sources_crosstalk_Linowski_2023,hypothesis_testing_Schlichtholz_2024,Boucher:20,coronagraphs_perfect_Deshler_experiment_2024}. From the plots, we can see that high detection probabilities corresponding to practically viable thresholds for false alarm probability require sample numbers that are large, but already attained in proof-of-principle experiments \cite{SPADE_superresolution_experiment_Rouviere_2024}. Furthermore, significant improvement in efficiency is visibly obtained by reducing the combined strength of crosstalk and dark counts, once again stressing the importance of noise mitigation.

\section{Example: 51 Pegasi b-like exoplanet}
\label{sec:Pegasi}
As an explicit example, let us consider the time $t$ needed to detect a Hot Jupiter-like exoplanet for set detection probability $P_\textnormal{det}$. Specifically, we choose a planet resembling 51 Pegasi b, the first discovery of an exoplanet orbiting a solar-type star \cite{first_solar-like_exoplanet_Mayor_1995}, as mentioned in the Introduction. The relevant system parameters include the absolute magnitude of the star $M\approx4.5$, the relative brightness $\nu\approx6 \cdot10^{-5}$, and the physical distance between the exoplanet and the star $d_{\textnormal{ex-st}}\approx0.05$~au \cite{first_solar-like_exoplanet_Mayor_1995,51pegasi_bright,51pegasi_dist}. As a reference telescope, we choose the Hubble Space Telescope with radius $r=1.2$~m \cite{esa_hubble_sm4_presskit_2009}, and consider a measurement in the visual light band with central frequency $\lambda=550$~nm, filter bandwidth $\Delta\lambda/\lambda=1\%$ and zero-point flux $F_{\lambda,0}\approx3.6\cdot10^{-2}\,\textnormal{W·m}^{-3}$ \cite{zero_point_flux_data_Bessell_1998}. 

From there, the ratio $s$ of the distance in the image plane to the width of the point spread function for a given distance from the telescope can be calculated as 
\begin{align}
    s=\frac{2rd_{\textnormal{ex-st}}}{0.62\lambda d_{\textnormal{sys}}}
        \approx \frac{1.71\,\text{pc}}{d_{\textnormal{sys}}},
\end{align}
where $d_{\textnormal{sys}}$ is the physical distance from the telescope to the probed system in parsecs. This is because the angular width of the point spread function equals approximately $0.62\lambda/2r$, while the angular separation is given by $d_{\textnormal{ex-st}}/d_{\textnormal{sys}}$. As for the sample size, the average number of photons collected by the telescope per unit time equals approximately
\begin{equation} \label{eq:N_Pegasi}
   \frac{N}{t} = (F_{\lambda}\Delta\lambda)\left(\pi r^2\right)\frac{\lambda}{h c},
\end{equation}
where $c$ is the speed of light, $h$ is the Planck constant, and the flux $F_{\lambda}$ can be calculated from the absolute magnitude of the star and its distance from the telescope using the standard formula $F_{\lambda}=F_{\lambda,0}10^{-0.4(M+5\log_{10}(d_{\textnormal{sys}})-5)}$ \cite{djorgovski_fluxes_2005}. To see that Eq. (\ref{eq:N_Pegasi}) is correct, note that its r.h.s. is just the incoming photon density times the area of the telescope divided by the energy of a single photon.

\begin{figure}
    \centering
    \includegraphics[width=\linewidth]{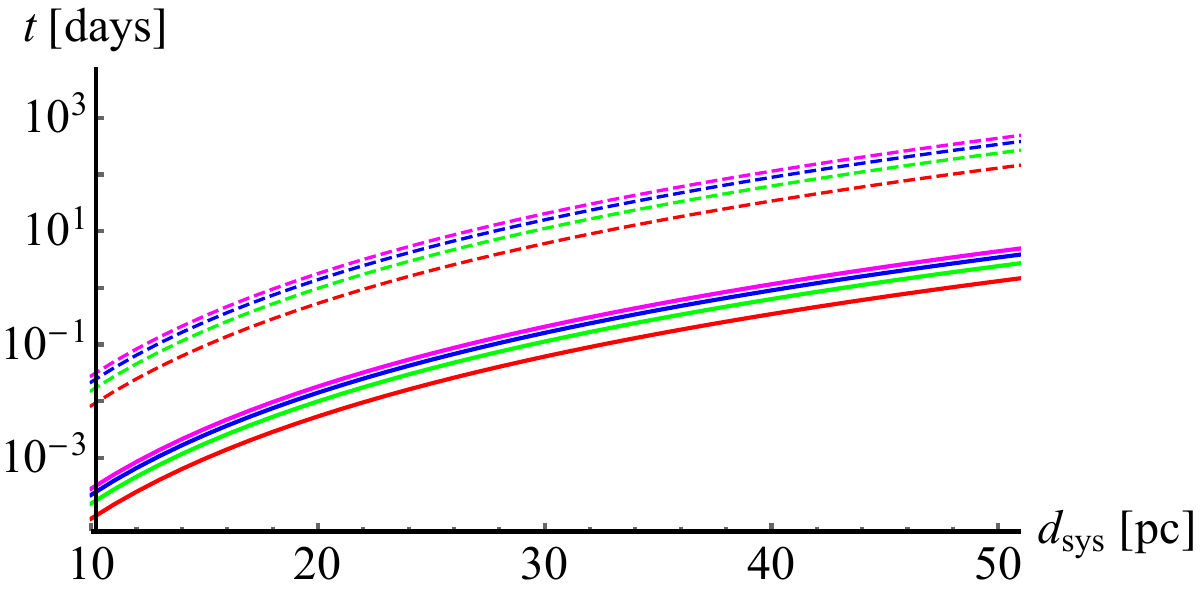}
    \caption{Detection time $t$ as a function of distance $d_{\textnormal{sys}}$ of the telescope from 51 Pegasi-like system with Hot Jupiter-like exoplanet (for 51 Pegasi itself, $d_{\textnormal{sys}}\approx 15$~pc). The two sets of curves, dashed and solid, correspond to large and small noise strengths, $p_0=10^{-2}$ and $p_0=10^{-4}$, respectively. Within each set, the different curves correspond to different target detection probabilities, from top to bottom: $P_\textnormal{det}=\{0.9999,\,0.999,\,0.99,\,0.9\}$ (magenta, blue, green, red). The threshold false alarm probability was set to $\bar{\alpha}=0.05$ in all the plots. As seen, the plotted detection times remain feasible in the considered parameter ranges.}
    \label{fig:time}
\end{figure}

Finally, using the last two equations, we can numerically invert Eq. (\ref{eq:beta_explicit}) to obtain the detection time $t$ as a function of desired detection probability $P_{\textnormal{det}}$, noise strength $p_0$, threshold false alarm probability $\bar{\alpha}$ and distance from the telescope $d_{\textnormal{sys}}$. In Fig. \ref{fig:time},
we plot the detection time obtained in this way as a function of $d_{\textnormal{sys}}$ within the range of our stellar neighborhood, for several representative values of the remaining parameters. As seen, most exoplanets in the range of tens of parsecs from the telescope, can be reliably detected in a matter of days, even for large detector noise. An exoplanet the same distance from the Earth as 51 Pegasi b, $d_{\textnormal{sys}} \approx 15$~pc, would take from a little over a~minute for small noise strengths up to less than seven hours for significantly noisy detectors.

\section{Conclusion and outlook}
\label{sec:summary}
We 
analyzed the practical efficiency of noisy SPADE in the context of asymmetric hypothesis testing for exoplanet detection:
First, we computed its detection probability and derived the explicit form of the statistical test maximizing it.
Then, we assessed its performance for planet-star system and detector noise values corresponding to realistic experimental conditions, and compared it with that of commonly used measurements such as direct imaging and coronagraphs.
Nearly all of our findings were derived analytically, allowing for an intuitive and precise understanding of their implications, including the explicit impact of noise mitigation on measurement efficiency. 

We found that, due to the vast difference in brightnesses between the planet and the star, the impact of experimental imperfections is much more detrimental than in related problems \cite{crosstalk_original_PRL,hypothesis_testing_Schlichtholz_2024}. 
Even so, we showed that while 
noisy SPADE has the same qualitative scaling with separation and relative brightness as direct imaging and coronagraphs, it outperforms them for realistic noise levels.
Accordingly, given the rapidly increasing performance of SPADE detection systems \cite{MPLC_2021,MPLC_2023,MPLC_2025}, we expect them to gain a prominent role in exoplanet detection in the near future.

Our findings not only mark an important step
for converting the theoretically
quantum-optimal ideal SPADE measurement into a practical technology, usable under realistic experimental conditions, but also suggest
several directions for future research. First, our results could be generalized to more sources, e.g., a star and multiple exoplanets, i.e., a multiple hypotheses testing problem. Furthermore, for particularly faraway solar systems, the expected integration times required to obtain a sufficient sample size could be so long that the natural motion of the planet and the star would impact the probability distributions \cite{Schlichtholz_2024_SPADE_DYNAMIC}, requiring accordingly tailored decision strategy. Similarly, the object planes associated with the star and the exoplanet may be different enough that one would have to measure the full three-dimensional separation vector, rather than only its length \cite{imaging_separation_vector_Yu_2018,imaging_separation_vector_Napoli_2019}. Finally, the observed significant impact of experimental imperfections on SPADE motivates the search for either novel noise mitigation techniques for SPADE, or alternatives to SPADE, which would be more robust to noise while still close to quantum-optimal. In this regard, one promising direction is to consider multi-photon detection events, which, while rare, can significantly help improve imaging resolution \cite{multi-photon_imaging_Katamadze_2025}, also in the context of astrometry \cite{multi-photon_imaging_astrometry_Chen_2023}.


\section*{Acknowledgements}
This work is carried out under IRA Programme, project no. FENG.02.01-IP.05-0006/23, financed by the FENG program 2021-2027, Priority FENG.02, Measure FENG.02.01., with the support of the FNP. This work was supported by the Fraunhofer Internal Programs under Grant No. Attract 40-09467.

\bibliography{report}{}
\bibliographystyle{obib}

\end{document}